\documentclass[showkeys,showpacs,prb,twocolumn,superbib]{revtex4-1}
\usepackage{amsmath}
\usepackage{color}
\usepackage{graphicx}
\usepackage{setspace}
\usepackage{tikz}

\newcommand{\sbs}{{Sb$_2$S$_3$}}
\newcommand{\bis}{{Bi$_2$S$_3$}}
\newcommand{\sbte}{{Sb$_2$Te$_3$}}
\newcommand{\gru}{{Gr\"uneisen}}
\newcommand{\invK}{K$^{-1}$}
\newcommand{\invcm}{cm$^{-1}$}

\newcommand{\VEC}[1]{{\boldsymbol #1}}

\begin{document}
\title{
Large anharmonic effect and thermal expansion anisotropy of metal chalcogenides: The case of antimony sulfide
}

\author{Chee Kwan Gan}
\email{Corresponding author: ganck@ihpc.a-star.edu.sg}
\affiliation{Institute of High Performance Computing, 1 Fusionopolis Way, \#16-16 Connexis, Singapore 138632}
\author{Jian Rui Soh}
\affiliation{Institute of High Performance Computing, 1 Fusionopolis Way, \#16-16 Connexis, Singapore 138632}
\author{Yun Liu}
\affiliation{Department of Materials Science and Engineering,
Massachusetts Institute of Technology,
77 Massachusetts Avenue,
Cambridge, MA 02139}

\date{Phys. Rev. B 92 (2015) 235202. Received 10 September 2015; published 2 December 2015}

\begin{abstract}
We derive a compact matrix expression for the linear
thermal expansion coefficients (TECs) for a general orthorhombic system
which relates elastic properties and integrated quantities
based on deformation and mode dependent \gru\ parameters and mode
dependent heat capacities. 
The density
of \gru\ parameters $\Gamma(\nu)$ as a function of frequency $\nu$,
weighted by the number of phonon modes, is introduced and found to be
illuminating in interpreting the TEC results. 
Using density-functional perturbation theory and \gru\ formalism
for thermal expansion, we illustrate the general usefulness of this method by  calculating the linear and volumetric thermal
expansion coefficients (TECs) of a low-symmetry orthorhombic compound antimony
sulfide (\sbs),
a compound belonging to a large class
of technologically and fundamentally important materials. Even though negative \gru\
parameters are found for deformations in all three crystal directions,
the $\Gamma(\nu)$ data rule out the occurrences of negative TECs
at all temperatures.  \sbs\ exhibits a large thermal expansion anisotropy 
where the TEC in the $b$ direction can reach as high as $13\times 10^{-6}$~\invK\ 
at high temperatures, about two and seven times larger than the TECs 
in the $c$ and $a$ direction, respectively.  Our work suggests a general
and practical first-principles approach to calculate the thermal
properties of other complicated low-symmetry systems.
\end{abstract}


\maketitle

\section{Introduction}

Metal chalcogenides comprise an important class of semiconductors
for optoelectronics,
photovoltaics, and thermoelectrics.\cite{Roy78v25,Deshmukh94v141,Porat88v87,Caracas05v32,Zhao11v84}
Recently, Raman spectroscopies and pump probe experiments
on examples such as \bis\ and \sbs\ have 
demonstrated the importance of phonons in modulating fundamental 
scattering processes\cite{Zhao11v84,Chong14v90}. 
Even though phonon dispersions have been reliably
obtained for some of 
these materials\cite{Zhao11v84,Liu14v16}, 
a first-principles study of the anharmonic effects
due to phonon-phonon scatterings that account for
thermal conductivities and thermal
expansion coefficients (TECs) has been lacking. This could be attributed to
the fact that metal chalcogenides 
have a relatively large primitive cell and 
a low-symmetry orthorhombic structure with three lattice
parameters, in contrast to some of the well-studied cubic structures with
a single lattice parameter\cite{Pavone93v48,Bohnen09v80}. 
TECs may routinely be calculated
using a direct minimization approach within the quasi-harmonic
approximation\cite{Mounet05v71}. 
However, for metal chalcogenides,
huge computational costs are needed
to perform many phonon calculations to locate a free energy
minimum at a given temperature in the three-dimensional lattice parameter
space.
Moreover, even if a direct minimization approach could be carried out,
it may be difficult to understand the underlying physics without 
investigating 
fundamental quantities such the \gru\ parameters, elastic constants,
heat capacities and mean-square displacements.

Here we should
mention a recent first-principles approach that is based on
the vibrational self-consistent-field
to calculate TECs\cite{Monserrat13v87}.
In another work\cite{Jiang09v80}, a nonequilibrium Green's function method
is used to calculate TECs of carbon nanotubes and graphene with a force-field potential.

\section{Methodology}

In this paper we adopt
the \gru\ formalism\cite{Gruneisen26v10,Barron80v29,Schelling03v68}
to predict the thermal properties of a low-symmetry orthorhombic system
with a first-principles method.
To the best of our knowledge it is the first time 
a first-principles thermal expansion study has been done on 
a crystal that is characterized by three lattice parameters.
We find that the linear TECs of an 
orthorhombic system in the $a$, $b$, and $c$ directions, denoted by
$\alpha_1$, $\alpha_2$, and $\alpha_3$, respectively,
at a temperature $T$ may be 
described by a matrix equation
\begin{equation}
\alpha = \frac{1}{\Omega}C^{-1} I,
\end{equation} 
where $ \alpha^T = (\alpha_1,\alpha_2, \alpha_3)$, 
$\Omega$ is the equilibrium volume of the primitive cell, and
$C^{-1}$ is the elastic compliance matrix\cite{Kittel96} with
matrix elements $C_{ij}$ being the elastic constants. 
A component $I_i(T)$ of the vector $I = (I_1, I_2, I_3)^T$ is given by 
$I_i(T) =  \frac{\Omega}{(2\pi)^3} \sum_{\lambda} \int_{\rm BZ} \gamma_{i,\lambda \VEC{k}} c(\nu_{\lambda\VEC{k}},T)\ d\VEC{k} $ where the 
integral is over the first Brillouin zone (BZ). 
A phonon mode with frequency $\nu_{\lambda \VEC{k}} $ is labeled by a mode index $\lambda$ and a wave vector $\VEC{k}$. 
The heat capacity of a phonon mode with frequency $\nu$ at temperature $T$ is
$c(\nu,T) = k_B  r^2/\sinh^2 r$, 
with $r = h\nu/2k_B T$. $h$ and $k_B$ are the 
Planck and Boltzmann constants, respectively.
The mode \gru\ parameters
$ \gamma_{i,\lambda \VEC{k}} = -\nu^{-1}_{\lambda \VEC{k}} \partial \nu_{\lambda \VEC{k}}/\partial \epsilon_i$ 
measure the relative change of phonon frequencies
$\nu_{\lambda \VEC{k}}$ as a result of deformations applied to the
crystal characterized by strain parameters $\epsilon_i$.

To illustrate the usefulness of our method,
we carry out density-functional theory calculations 
using the plane-wave basis Quantum Espresso suite\cite{Giannozzi09v21} 
on \sbs, an example of metal chalcogenides with a 
small direct bandgap of $1.5$~eV.
The local density approximation is used to describe the exchange-correlation. Pseudopotentials based on the
Rappe-Rabe-Kaxiras-Joannopoulos\cite{Rappe90v41} approach as found in the ``atomic code'' of the standard Quantum Espresso distribution
are used.
A large cutoff energy of $60$~Ry is used throughout and a Monkhorst-Pack mesh of $4\times 12 \times 4 $ is used 
for $k$-point sampling. 
Atomic relaxation is stopped when the forces on all the atoms are less than $1$~meV/\AA.
We use the nonsymmorphic space group $Pnma$ to describe \sbs\ with 20 atoms in a primitive cell, of which 5 are inequivalent. 
We obtain $a_0 = 11.021$, $b_0 = 3.797$~, and $c_0 = 10.783$~\AA, in good agreement with experimental values.\cite{Lundegaard03v30}
The phonon modes are calculated using the density-functional perturbation theory.\cite{Baroni01v73}
Phonon calculations are carried out on a $q$ mesh of $2\times 4 \times 2$, which is
equivalent to a $2\times 4\times 2$ supercell force-constant\cite{Gan06v73} phonon calculation, the efficacy of which
has been confirmed\cite{Liu14v16}. We note that the results do not appreciably change when we use a larger $q$ mesh of $3\times 6 \times 3$.
For the $q$ mesh of $2\times 4 \times 2 $,
dynamical matrices have to be calculated at
12 irreducible $q$ points. For a general $q$ point, 
one has to loop through 60 irreducible representations, each of which
requires a number of self-consistent-field calculations. Interestingly, the seemingly high-symmetry point $\Gamma$ 
has a relatively large number of irreducible representations of $60$, which incurs more 
computation costs compared to, say, a diamond crystal with only 2 irreducible representations at $\Gamma$. The 12-$q$ points
correspond to a total of 471 irreducible representations.
Ignoring the cost for convergence tests, we already need to handle a minimum of $7 \times 471 = 3297$ irreducible representations to 
carry out a central-difference scheme for $a$, $b$, and $c$ directions (note that we need to
perform a set of phonon calculations on the equilibrium structure). The cost analysis
also suggests
even larger computational resources will be 
required if one wishes to carry out a full direct minimization study based on the 
quasi-harmonic approximation in finding the free energy minimum at each temperature in the three-dimensional search
space of $\{a,b,c\}$.
At the end of self-consistent-consistent calculations, all
dynamical matrices are collected and 
interatomic force constants are obtained by an inverse Fourier transform. The Brillouin zone sampling for the 
integrated quantities $I_i(T)$ is over a large mesh of $15\times 45 \times 15$.
We perform standard elastic constants calculations\cite{Beckstein01v63,Koc12v14}
to obtain $(C_{11}$, $C_{12}$, $C_{13}$, $C_{22}$, $C_{23}$, $C_{33}) = (133.19$, $36.45$, 
$55.99$, $141.08$, $67.14$, $119.09)$~GPa.

\begin{figure}[htbp]
\centering
\includegraphics[width=8.0 cm,clip]{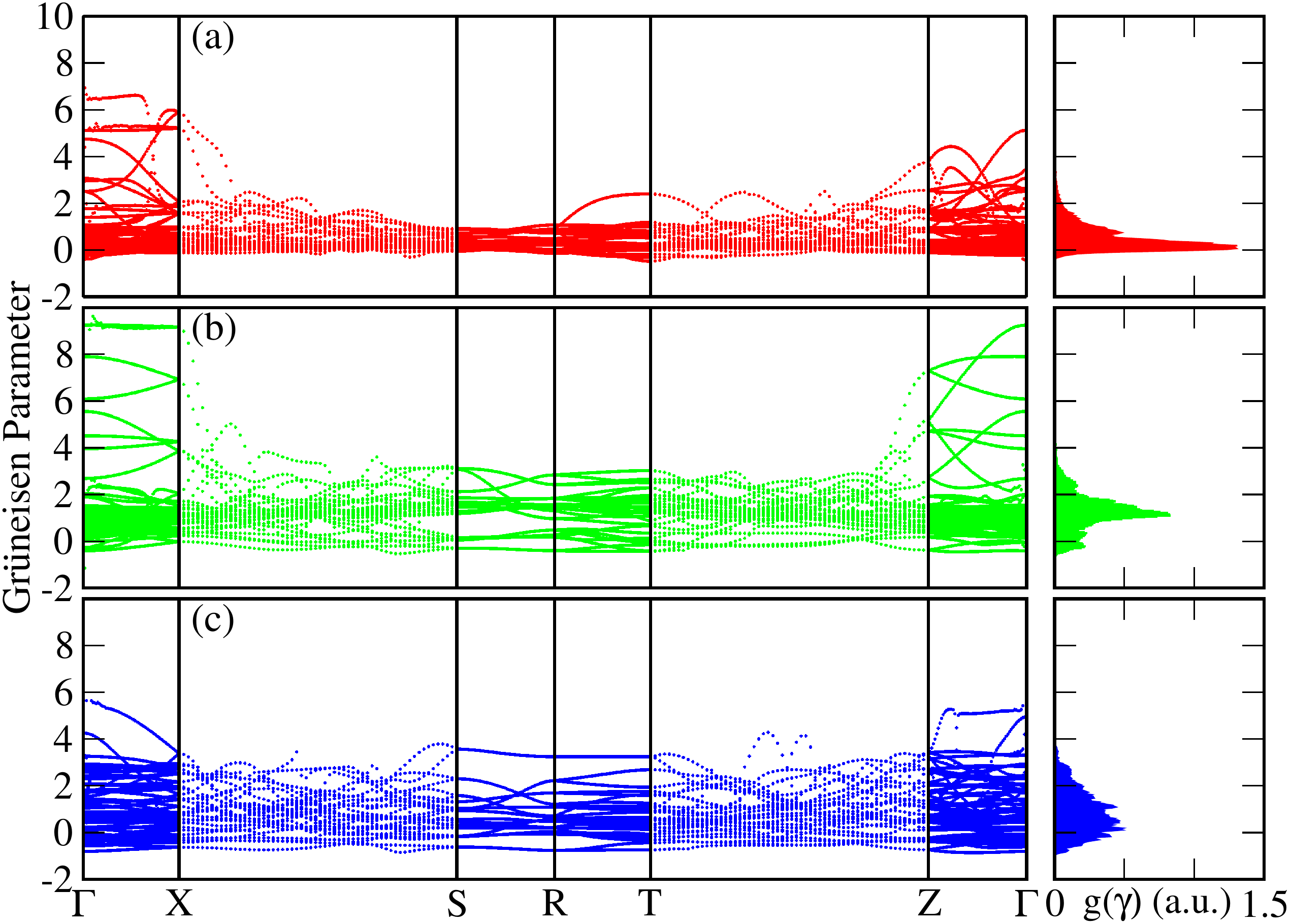}
\caption{(Color online) \gru\  parameters $\gamma_{i,\lambda \VEC{k}}$ along the high-symmetry directions for orthorhombic \sbs,
for $i= 1$, $2$, and $3$, corresponding to deformations
due to strains $e_1$, $e_2$, and $e_3$, are shown 
in (a), (b), and (c), respectively. The coordinates of $X$,
$S$, $R$, $T$, and $Z$ can be found in Ref.~[\onlinecite{Liu14v16}].
The corresponding plain densities of \gru\ parameters are shown on the right.
}
\label{fig:gru-para}
\end{figure}

\section{Results and discussions}

\begin{figure}[htbp]
\centering
\includegraphics[width=8.0 cm,clip]{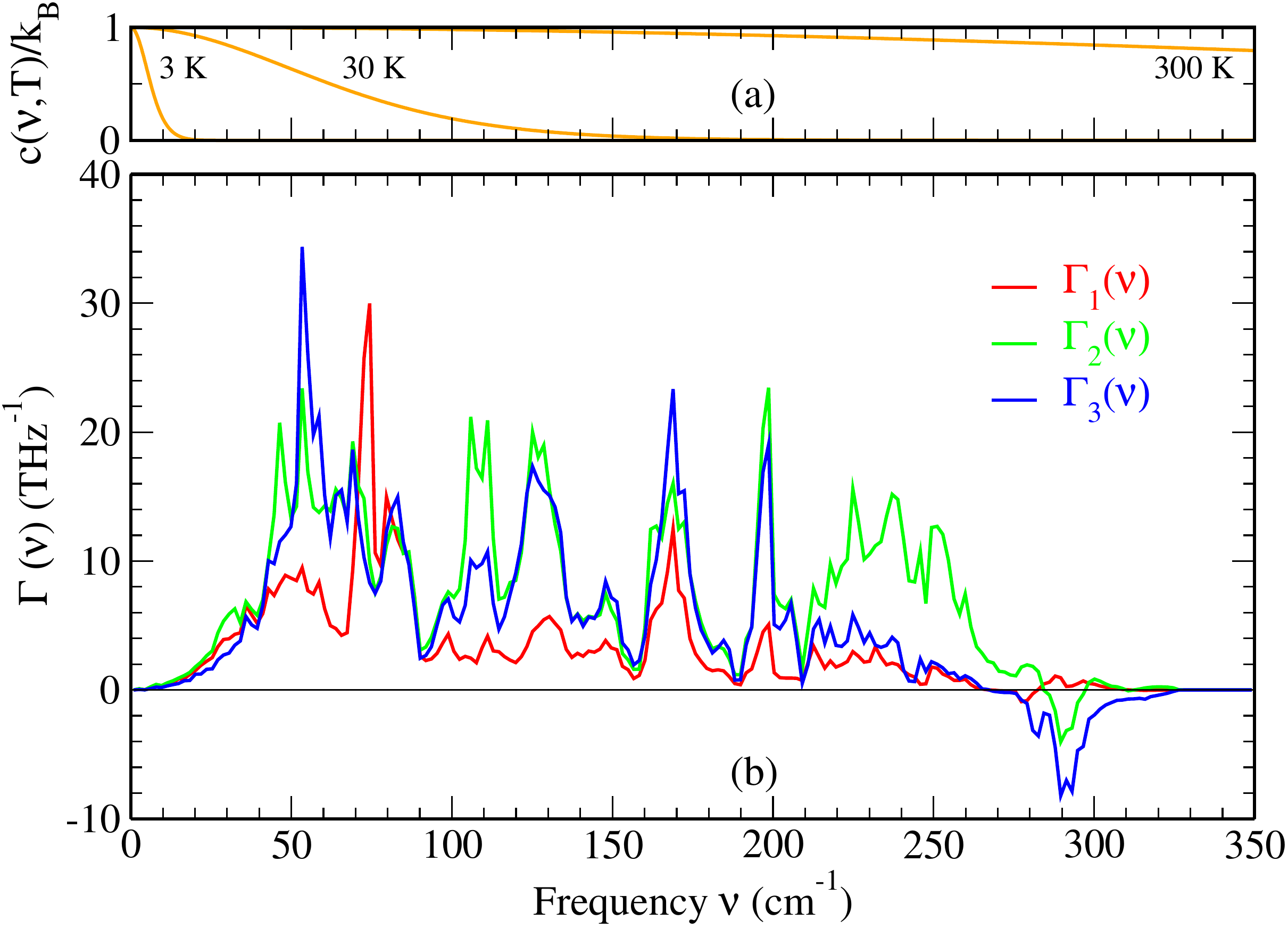}
\caption{(Color online) (a) The mode dependent heat capacity as a function of phonon frequency $\nu$ for three representative temperatures.
(b) Density of \gru\ parameters.
}
\label{fig:density}
\end{figure}

We apply strains of $\epsilon= \pm 0.005$ (strains of $\epsilon = \pm 0.010$ do not change the results appreciably)
to obtain the central-difference \gru\ parameters 
in the $a$, $b$ and $c$ directions. By using the change in the dynamical matrix
resulting from a finite deformation to the crystal and with the help of first-order perturbation theory, 
we can determine the change of frequency for each phonon mode to 
obtain the \gru\ parameters. The results of $\gamma_{i,\lambda\VEC{k}}$ are shown in 
Fig.~\ref{fig:gru-para}. It is noticed that along the high-symmetry directions, the
degeneracy of the \gru\ parameters is preserved.
There are some bands that have large values (say, $> 4$) of \gru\ parameters. 
By performing a $k$-point sampling over the BZ, we
calculate the plain density of \gru\ parameters
$g_i (\gamma) = \frac{\Omega}{(2\pi)^3} \sum_{\lambda} \int_{\rm BZ} \delta(\gamma - \gamma_{i,\lambda \VEC{k}}) \  d\VEC{k}$, results 
of which are shown on
the right panels of Fig.~\ref{fig:gru-para}.
The $g_i(\gamma)$ plots show that
large \gru\ parameters are not highly populated. 
Interestingly, $g_i(\gamma)$ show some population
of negative \gru\ parameters, especially for $g_3(\gamma)$, which may lead to negative TECs if these 
negative parameters correspond to low-frequency modes.
In the literature, one can use the average \gru\ parameters\cite{Bohnen09v80} or the 
scattered $\gamma$-$\nu$ plot\cite{Stoffel15v27} to display this information. However, here we propose 
a quantity called the density of \gru\ parameters $\Gamma_i(\nu)$, weighted by the 
number of phonon modes, defined as $\Gamma_i (\nu) = \frac{\Omega}{(2\pi)^3} \sum_{\lambda} \int_{\rm BZ} 
\delta(\nu - \nu_{\lambda \VEC{k}}) \gamma_{i,\lambda \VEC{k}}\  d\VEC{k} $ to capture the collective effects of \gru\ parameters
and phonon frequencies.
Apart from its direct physical meaning, $\Gamma_i(\nu)$ also allows a second equivalent expression for $I_i (T)$, which 
is $\int_{\nu \rm min}^{\nu \rm max} \Gamma_i (\nu) c(\nu,T)\ d\nu$.
The densities of \gru\ parameters  $\Gamma_i(\nu)$ are
shown in Fig.~\ref{fig:density}(b), where the negative \gru\ parameters
are confined to phonon frequencies of around $290$~\invcm. These phonons are not excited at low 
temperatures (see Fig.~\ref{fig:density}(a) for the dependence of $c(\nu,T)$ on $\nu$ for 
representative temperatures of $3$, $30$ and $300$~K). 

\begin{figure}
\centering
\includegraphics[width=8.0 cm,clip]{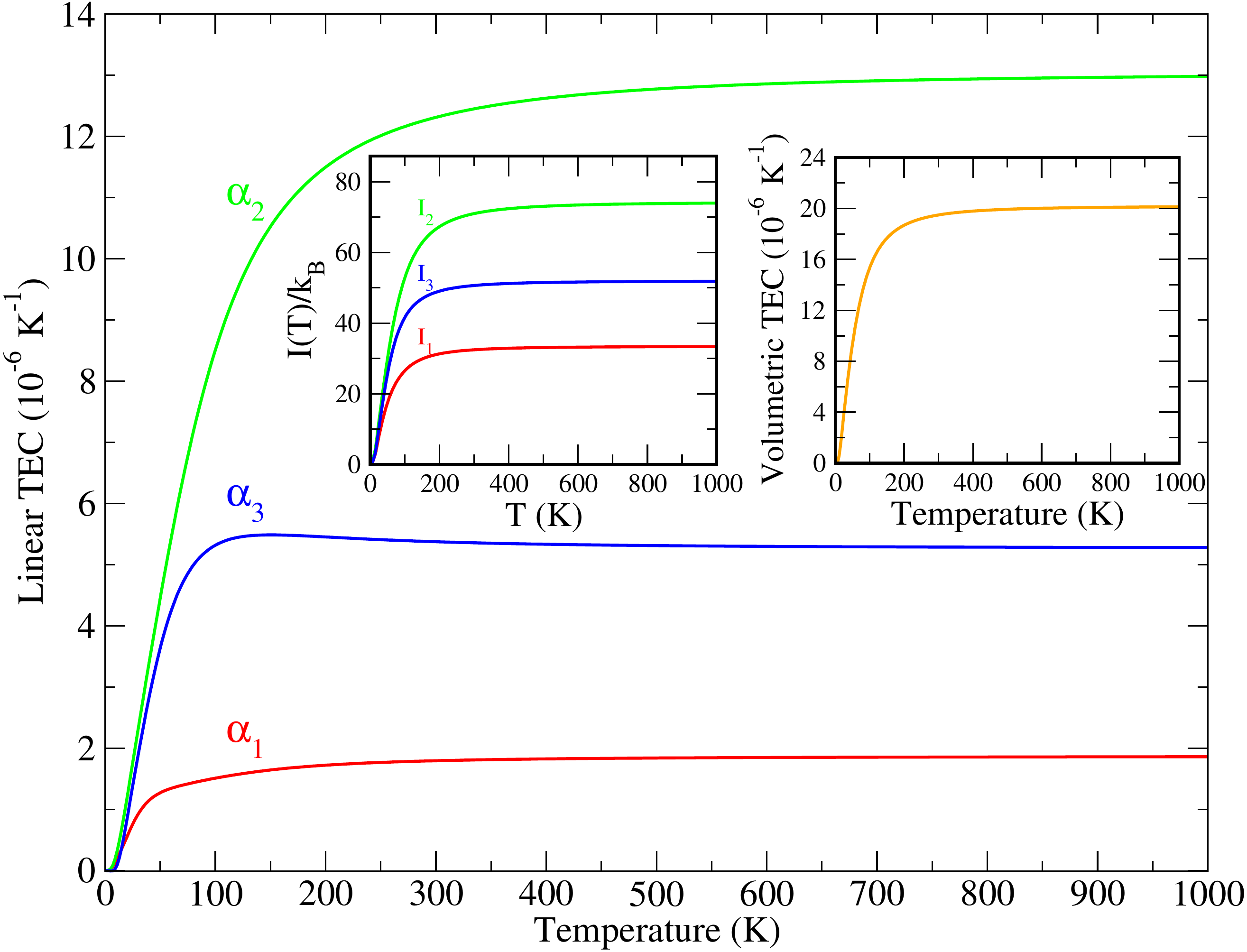}
\caption{(Color online) The linear TECs of \sbs\ as functions of temperature. The insets show the integrated quantities $I_i(T)$ and the volumetric TEC.
}
\label{fig:tec}
\end{figure}

\begin{figure}
\centering
\includegraphics[width=8.0 cm,clip]{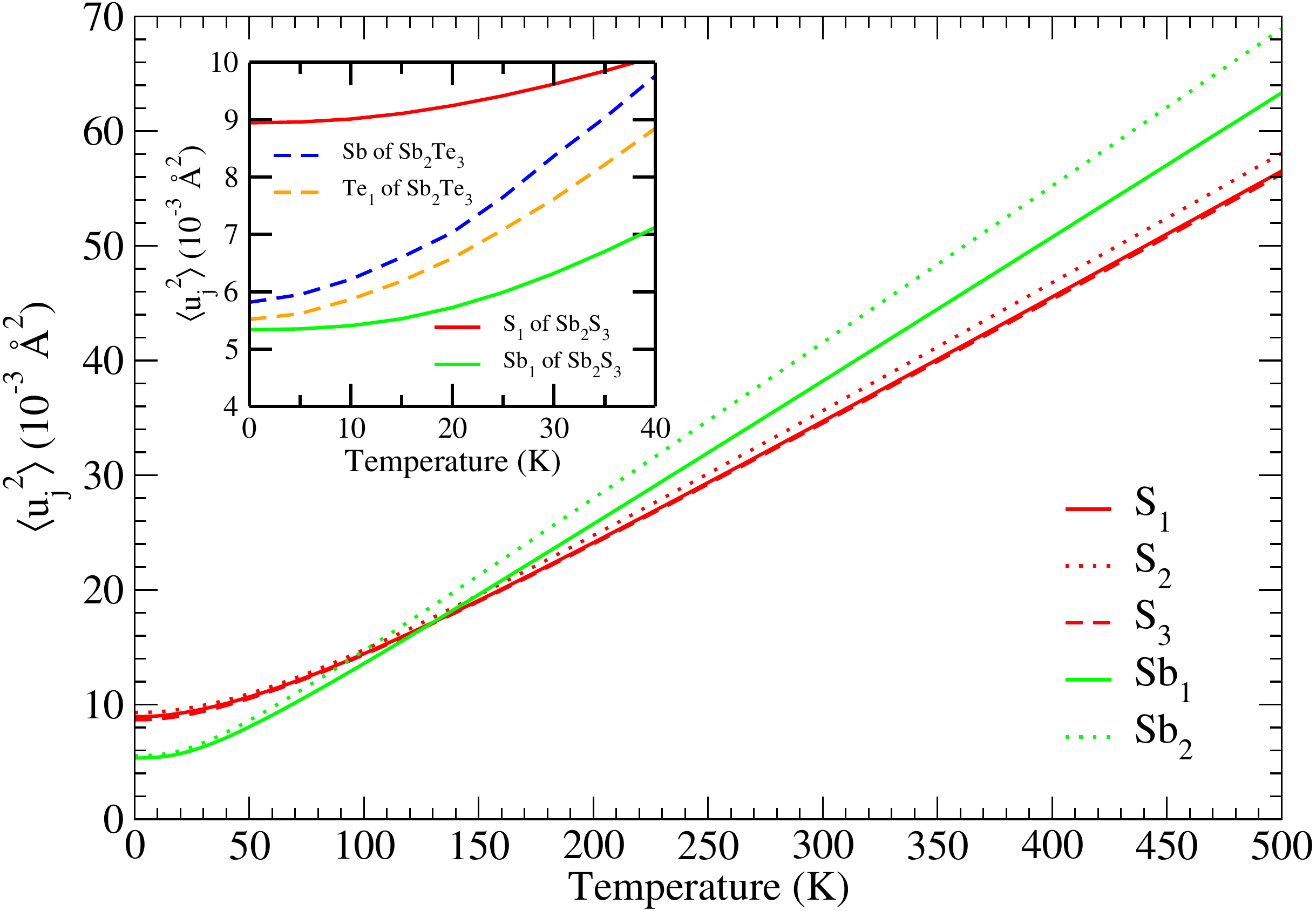}
\caption{(Color online) MSDs of five inequivalent atoms
of \sbs\ as functions of temperature. 
The inset shows the comparison of the MSDs of representative atoms in
\sbs\ and a reference trigonal
system of \sbte.
The MSD for the $j$th atom is\cite{Yang12v98} calculated from
$\langle
u_j^2 \rangle = 
\frac{\Omega}{(2\pi)^3} \sum_{\lambda} \int_{\rm BZ} \frac{ h |e_j(\lambda,\VEC{k})|^2 \coth r    }{ 8 \pi^2 M_j \nu_{\lambda\VEC{k}}      }\ d\VEC{k} $
with $r = h\nu_{\lambda\VEC{k}}/2k_B T$ and $e_j(\lambda,\VEC{k})$ 
the eigenvector for the $j$th atom of mass $M_j$.
}
\label{fig:msd}
\end{figure}

The results of the linear and volumetric thermal expansion coefficients are shown in Fig.~\ref{fig:tec}. 
Except for $\alpha_3$ beyond $150$~K, all TECs are monotonically increasing functions of temperature. 
$I_i(T)$ (shown in an inset of Fig.~\ref{fig:tec}) also exhibits a largely similar temperature dependence. 
Despite the occurrences of negative \gru\ parameters as shown in Fig.~\ref{fig:gru-para}, all
linear and volumetric TECs are positive.
At high temperatures, the effect of
phonon modes with negative \gru\ parameters is canceled out by the
more highly populated phonon modes at lower frequencies with
positive \gru\ parameters, thus eliminating the possibility of negative TECs at any temperature.
The high-temperature limits are
$\alpha_1 = 1.86$, $\alpha_2 = 13.0$, and $\alpha_3 = 5.28 \times 10^{-6}$~K$^{-1}$, and the volumetric TEC is 
$20.14 \times 10^{-6}$~K$^{-1}$. The small TEC of $\alpha_1$ among all other TECs is consistent with
the fact that $I_1(T)$ is smaller than $I_2(T)$ and $I_3(T)$, in addition to the fact that 
$C_{11}$ is comparable to $C_{22}$ but larger than $C_{33}$.
To the best of our knowledge, 
the experimental TEC
data of \sbs\ are not readily available for a direct comparison. 
However, we note that 
Stoffel {\it et al.}\cite{Stoffel15v27}
have demonstrated
that the volumetric TEC of trigonal \sbte\ using a quasi-harmonic 
approximation is accurate up to 300~K where the mean-square displacements (MSDs) of \sbte, shown
in the inset of Figure~\ref{fig:msd} compare rather well with
that of \sbs. Therefore
we believe our TEC results on \sbs\ are reasonable below 40~K. 
We note that the MSDs of \sbte\ at higher temperatures (which are not available in Ref.~[\onlinecite{Stoffel15v27}]) may
provide a better estimate on the temperature below which our TEC results are valid.
At low temperatures below $100$~K, Figure~\ref{fig:msd} shows
the three S atoms have larger MSDs than Sb, consistent
with the fact that S has a smaller mass compared to Sb. 
At high temperatures, however, the MSDs of Sb atoms are larger than that of the S atoms. 

\section{Summary}

In summary we have extended the \gru\ formalism to treat a low-symmetry structure of orthorhombic 
antimony sulfide. Using this approach, we applied just six deformations to the crystal to obtain the 
\gru\ parameters, thus avoiding the
huge computation requirement for a direct minimization based on
the quasi-harmonic approximation.
Even though negative \gru\ parameters were found, there are no negative TECs at all temperatures since these parameters
are associated with high frequency modes at around $290$~\invcm. 
It is expected that a similar approach could be used to 
address TECs of other low-symmetric systems such as those with monoclinic crystal structure.

\section{Acknowledgments}

C.K.G. acknowledges fruitful discussions with Aloysius Soon, Mark H. Jhon, and Hwee Kuan Lee.
J.R.S. and Y.L. acknowledge support from the Singapore National Science Scholarship.
We thank the A*STAR Computational Resource Center for computing resources.

\end{document}